\documentclass[prl,twocolumn,aps,letterpaper, preprintnumbers,superscriptaddress]{revtex4}
\usepackage{hyperref}
\usepackage{verbatim}
\usepackage{amsmath}
\usepackage{latexsym}
\usepackage{revsymb}
\usepackage{yfonts}
\usepackage{ifthen}
\usepackage{tcolorbox}
\usepackage{natbib}
\usepackage{amsfonts}
\usepackage{amsmath}
\usepackage{amssymb}
\usepackage{amsthm}
\usepackage{graphicx}
\usepackage{bm}
\usepackage{bbm}
\usepackage{epsfig,color,amssymb}
\usepackage{subfigure}
\usepackage{amsfonts}
\usepackage{amscd}
\usepackage{amsmath}
\usepackage{multirow}
\usepackage{chemarrow}
\usepackage{dcolumn}
\usepackage{bm}
\usepackage{graphicx}
\usepackage{enumerate}
\usepackage{epsfig}
\usepackage{subfigure}
\usepackage{xcolor}
\usepackage{multirow}
\usepackage{ulem}
\usepackage{upgreek}
\usepackage{braket}
\usepackage{comment}
\usepackage{enumitem}
\usepackage{amsthm}
\usepackage{color,soul}

\renewcommand{\emph}[1]{\textit{#1}}

\begin{document}

\title{Topological Transformation and Free-Space Transport of Photonic Hopfions}

\author{Yijie Shen}\email{y.shen@soton.ac.uk}
\affiliation{Optoelectronics Research Centre \& Centre for Photonic Metamaterials, University of Southampton, Southampton SO17 1BJ, United Kingdom}
\author{Bingshi Yu}
\author{Haijun Wu}
\author{Chunyu Li}
\author{Zhihan Zhu}\email{zhuzhihan@hrbust.edu.cn}
\affiliation{Wang Da-Heng Center, Heilongjiang Key Laboratory of Quantum Control, Harbin University of Science and Technology, Harbin 150080, China}
\author{Anatoly V. Zayats}\email{a.zayats@kcl.ac.uk}
\affiliation{Department of Physics and London Centre for Nanotechnology, King’s College London, London WC2R 2LS, United Kingdom}
\date{\today}


\begin{abstract}
Structured light fields embody strong spatial variations of polarisation, phase and amplitude. Understanding, characterization and exploitation of such fields can be achieved through their topological properties. Three-dimensional (3D) topological solitons, such as hopfions, are 3D localized continuous field configurations with nontrivial particle-like structures, that exhibit a host of important topologically protected properties. Here, we propose and demonstrate photonic counterparts of hopfions with exact characteristics of Hopf fibration, Hopf index, and Hopf mapping from real-space vector beams to homotopic hyperspheres representing polarisation states. We experimentally generate photonic hopfions with on-demand high-order Hopf indices and independently controlled topological textures, including N\'eel-, Bloch-, and anti-skyrmionic types. We also demonstrate a robust free-space transport of photonic hopfions, thus, showing potential of hopfions for developing optical topological informatics and communications.
\end{abstract}

\maketitle

Topological solitons with topologically protected spin texture are of fundamental interest in exploring fascinating physical phenomena and nonlinear field theories~\cite{manton2004topological}. In particular, numerous low-dimensional topological solitonic textures have been extensively studied in the recent years, such as 1D magnetic domain walls and magnetic skyrmions~\cite{parkin2008magnetic,wang2012domain,ryu2013chiral,han2019mutual,jue2016chiral}, as well as 2D electromagnetic skyrmions~\cite{tsesses2018optical,du2019deep,davis2020ultrafast,shen2021supertoroidal}, merons and bimerons~\cite{jani2021antiferromagnetic,dai2020plasmonic,shen2021topological}. Their sophisticated topological structures have been considered as promising information carriers in the next generation of data storage and communication devices~\cite{shen2022generation,gobel2021beyond,fert2017magnetic,bernevig2022progress}. 3D topological textures have also been attracting considerable interest in condensed matter physics and optics, following their theoretical introduction, extending the frontiers of topological field manipulation~\cite{gobel2021beyond,ackerman2017static,tang2021magnetic,Zheng2021}. 

Hopfions, as the most classic 3D topological solitons, were initially proposed in the Skyrme-Faddeev model~\cite{faddeev1976some,skyrme1991non,skyrme1962unified}. They are also known as Faddeev-Hopf knots~\cite{hietarinta1999faddeev,sutcliffe2007knots} and can be elegantly mapped to a Hopf fibration and characterized by a Hopf index~\cite{lyons2003elementary,whitehead1947expression}. Hopfions have fundamental importance in many physical systems in high-energy physics~\cite{faddeev1997stable}, chiral magnets~\cite{tai2018static,wang2019current,liu2020three,luk2020hopfions,gobel2020topological,kent2021creation}, quantum fields~\cite{hall2016tying,lee2018synthetic}, condensed matter physics~\cite{babaev2002dual,kawaguchi2008knots,bisset2015robust,zou2021formation}, cosmology~\cite{cruz2007cosmic,thompson2015classification}, fluid dynamics~\cite{kleckner2013creation}, liquid crystals~\cite{ackerman2017diversity,ackerman2015self}, and very recently realized in free-space photonics~\cite{sugic2021particle}. Due to their rich 3D spin texture, hopfions can potentially provide many opportunities in investigations and applications of topological structures, however, only very limited number of hophions were experimentally realized to date. For instance, magnetic hopfions of Bloch- (vortex) and N\'eel- (hedgehog) types can be excited and exist in a stable state in chiral magnets~\cite{wang2019current}. In photonics, only fundamental-order hopfions with a unit Hopf index were reported~\cite{sugic2021particle}. Generation and properties of higher-order hopfions and their topological spin texture tuning still need to be explored.

In this Letter, we theoretically and experimentally demonstrate a generalized family of photonic hopfions constructed using superposition of Laguerre-Gaussian (LG) beams with customized polarization patterns. The generated complex vector fields have spatially varying 3D polarisation distributions, which exhibits hopfion topological textures. The observed photonic hopfions can be freely transformed to various topological states, including N\'eel-type (hedgehog), Bloch-type (vortex), and anti-type (saddle) textures and many intermediate topological states. For each on-demand texture, a topological charge (Hopf index) of hopfions can also be tuned to arbitrary integers, realising higher-order hopfions. We also demonstrate a free-space propagation of hopfions with protected topology, revealing potential to develop technology of topological information transfer. 
\begin{figure*}[t!]
	\centering
	\includegraphics[width=\linewidth]{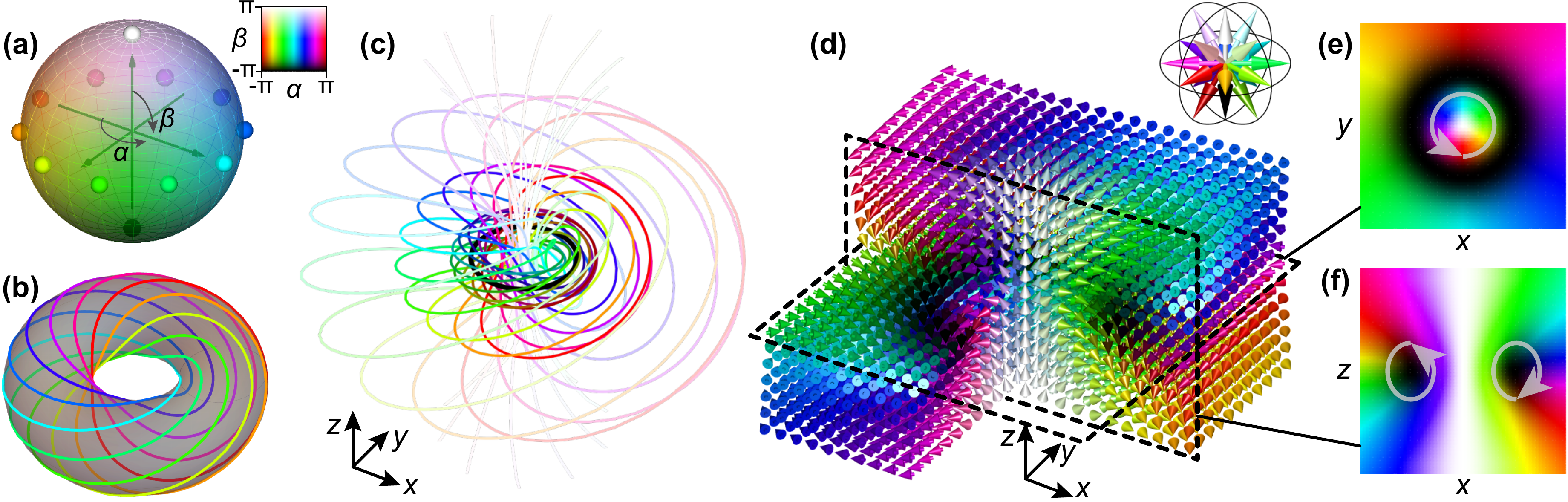}
	\caption{(a) The parameter-space visualization of a hypersphere: the longitude and latitude degrees ($\alpha$ and $\beta$) of a parametric 2-sphere are represented by hue color and its lightness (dark towards the south pole, where spin is down, and bright towards the north pole, where spin is up). Each point on a parametric 2-sphere corresponds to a closed iso-spin line located in a 3D Euclidean space. (b) The lines projected from the selected points of the same latitude $\beta$ and different longitude $\alpha$ on the hypersphere (highlighted by the solid dots with the corresponding hue colors), form torus knots covering a torus (with different tori corresponding to different $\beta$). (c) The real-space visualization of a Hopf fibration as a full stereographic mapping from a hypersphere: toroidally knotted lines (torus knots) arranged on a set of coaxially nested tori, with each torus corresponding to different latitude $\beta$ of a parametric 2-sphere. The black circle corresponds to south pole (spin down) and the axis of the nested tori corresponds to the north pole (spin up) in (a). (d) The 3D spin distribution in a hopfion, corresponding to the isospin contours in (c) with each spin vector colored by its $\alpha$ and $\beta$ parameters of a parametric sphere in (a) as shown in the insert. (e,f) The cross-sectional view of the spin distribution in (d): (e) $xy$ ($z=0$) and (f) $yz$ ($x=0$) cross-sections show skyrmion-like structures with the grey arrows marking the vorticity of the skyrmions. Colour scale is the same as correspond to the spin direction in (d).} 
	\label{f1}
\end{figure*}

In topology, a hopfion configuration is usually represented by a 3D real-space distribution of generalised unit spin vectors, $\mathbf{S}(x,y,z)=[s_x(x,y,z),s_y(x,y,z),s_z(x,y,z)]$, and achieved by a stereographic projection from a hypersphere $\mathbb{S}^3$ ($\bm{\chi}=(\chi_1,\chi_2,\chi_3,\chi_4)$) to a real space $\mathbb{R}^3$ ($\mathbf{r}=(x,y,z)$), fulfilling the Hopf map $\mathbf{S}=\langle\bm{\zeta}|\bm{\sigma}|\bm{\zeta}\rangle$, where $|\bm{\zeta}\rangle=(\chi_4+\text{i}\chi_3,\chi_1+\text{i}\chi_2)^{\text{T}}$ and $\bm{\sigma}=[\bm{\sigma}_x,\bm{\sigma}_y,\bm{\sigma}_z]$ corresponds to the three Pauli matrices~\cite{liu2020three}. As a geometric representation of a Hopf map, a point on a $\mathbb{S}^2$ sphere (a so-called 2-sphere) corresponds to a closed line (loop) in a real space rather than a point in a real space, revealing the additional dimension of a hypersphere. Each loop corresponds to an isospin contour ($(s_x, s_y, s_z)$= const) in the hopfion texture. Extending this description to optics, $\bm{\chi}$ components correspond to the real and imaginary parts of the complex-valued fields projected onto the circular polarisation basis, thus capturing both polarisation and phase variations in a 3D space. In turn, $\mathbf{S}$ is, therefore, formed by Stokes vectors $S_1$, $S_2$ and $S_3$, describing spatial variations of polarisation of an optical field, and a 2-sphere is equivalent to a Poincar\'e sphere. In this framework, the isospin contours are iso-polarisation contours of an optical vector field. 

The example of the mapping is shown in Fig.~\ref{f1}, with each color of a point on a parametric 2-sphere representing a spin vector with specific latitude and longitude, $\alpha$ and $\beta$, shown in Fig.~\ref{f1}(a) with hue color and lightness, respectively, and the corresponding stereographic projection of an iso-polarisation contour in a real space is shown as a loop with the same color (Fig.~\ref{f1}(b,c)). The loops mapped from the points on the same latitude $\beta$ of a parametric sphere form a set of the torus knots (a set of knots that lie on the surface of a special torus), completely covering a torus when scanning the points with different longitudes $\alpha$ (Fig.~\ref{f1}(b)). Thus, the full unwrapping of the parametric sphere (for all latitudes and longitudes) results in the torus knots on the nested tori with each torus corresponding to different latitude $\beta$, akin to matryoshka (Fig.~\ref{f1}(c)); this construction is termed the Hopf fibration. In two extreme cases of the mapping, the spin-down contour (left circular polarisation in optics), corresponding to the south pole of a 2-sphere, is the ring at the core of the nested tori (black line), while the spin-up contour (right circular polarisation), corresponding to the north pole, is the line in $z$-direction along the axis of the nested tori. 

How knotted each torus knot is depends on the topological charge (Hopf index, $Q_H$) {(see Supplementary Note I and Fig. S1)}~\cite{hilton1953introduction}. The Hopf index (topological charge) is determined by an around-torus index $p$ and a through-torus index $q$ via $Q_H = p\times q$ describing how many times an isospin contour goes around the torus ($p$) and through the torus hole ($q$). A fundamental hopfion corresponds to one around and one through a torus configuration ($Q_H$=1). 

The 3D hopfion texture is shown in Fig.~\ref{f1}(d), where the colors of vectors correspond to the isospin contours in (c). The 3D hopfion is closely related to 2D skyrmions. Hopfion textures have a skyrmionium (a doughnut-shaped skyrmion) in $xy$ cross-section at $z=0$ (Fig.~\ref{f1}(e)) and the $yz$ cross-section ($x=0$) corresponds to two skyrmions with opposite vorticity and opposite skyrmion topological charges (Fig.~\ref{f1}(f)). These are the signatures of a hopfion since the hopfion can be represented by an end-to-end twisted skyrmion tube~\cite{gobel2020topological}.  Due to the closed connection to skyrmions, hopfions can be classified by the type of skyrmionium in the $xy$ cross-section, which in turn is classified by the type of 2D skyrmions forming it {(see Supplementary Note II for details)}~\cite{shen2021topological,shen2022generation,gobel2021beyond}.

Hopfion is defined by~\cite{whitehead1947expression} $
Q_{H}= p\times q = \frac{1}{(4\pi)^2 }\iiint{\mathbf{F}\cdot\mathbf{A}}\text{d}x\text{d}y\text{d}z$, where $F_i=\varepsilon_{ijk}\mathbf{S}\cdot(\partial_j\mathbf{S}\times\partial_k\mathbf{S})/2$, in which $i,j,k=\{x,y,z\}$, $\varepsilon$ is the Levi-Civita tensor, and $\mathbf{A}$ is the vector potential satisfying $\nabla\times\mathbf{A}=\mathbf{F}$ and is closely related to the skyrmion number ($p$ and $q$ correspond to the skyrmion numbers of the skyrmioniums in the $xz$ and $xy$ cross-sections of a hopfion~\cite{ackerman2015self}), which can be defined by the product of topological numbers of polarity $Q_P$ and vorticity $Q_V$ of the vector field~\cite{gobel2021beyond,shen2022topological}. The polarity $Q_P=\frac{1}{2}[\cos\beta (r)]_{r=0}^{r=a}=\pm 1$ is defined by the vector direction down (up) at the center $r=0$ and up (down) at the skyrmion boundary $r\to r_\sigma$ for $Q_P=1$ ($Q_P=-1$), and the vorticity $Q_V=\frac{1}{2\pi }[\alpha (\phi )]_{\phi =0}^{\phi =2\pi }$ can be an arbitrary integer, which controls the azimuthal distribution of the transverse vector field components. For a fixed vorticity, an initial phase $\theta$ should be added to distinguish the helicity to completely decide the transverse distribution of the vector field, i.e. $\alpha(\phi)=m\phi+\theta$, which is termed helicity. For instance, skyrmion with vorticity charge $Q_V=1$ represents a hedgehog-like texture for a helicity $\theta=0$ (named N\'eel-type-I), a squeezed hedgehog-like texture when helicity is $\pi$ (named N\'eel-type-II), and a right- or left-handed vortex texture when helicity is $\pi/2$ or $3\pi/2$ (named Bloch-type). While, for the vorticity charge of $-1$, the skyrmion has a saddle-like texture (named anti-type). The classification can be applied to hopfions, i.e. N\'eel-, Bloch-, anti-type hopfions, given that the skyrmion appears in the $xy$ cross-section of the hopfion. 

\begin{figure}[t!]
	\centering
	\includegraphics[width=0.9\linewidth]{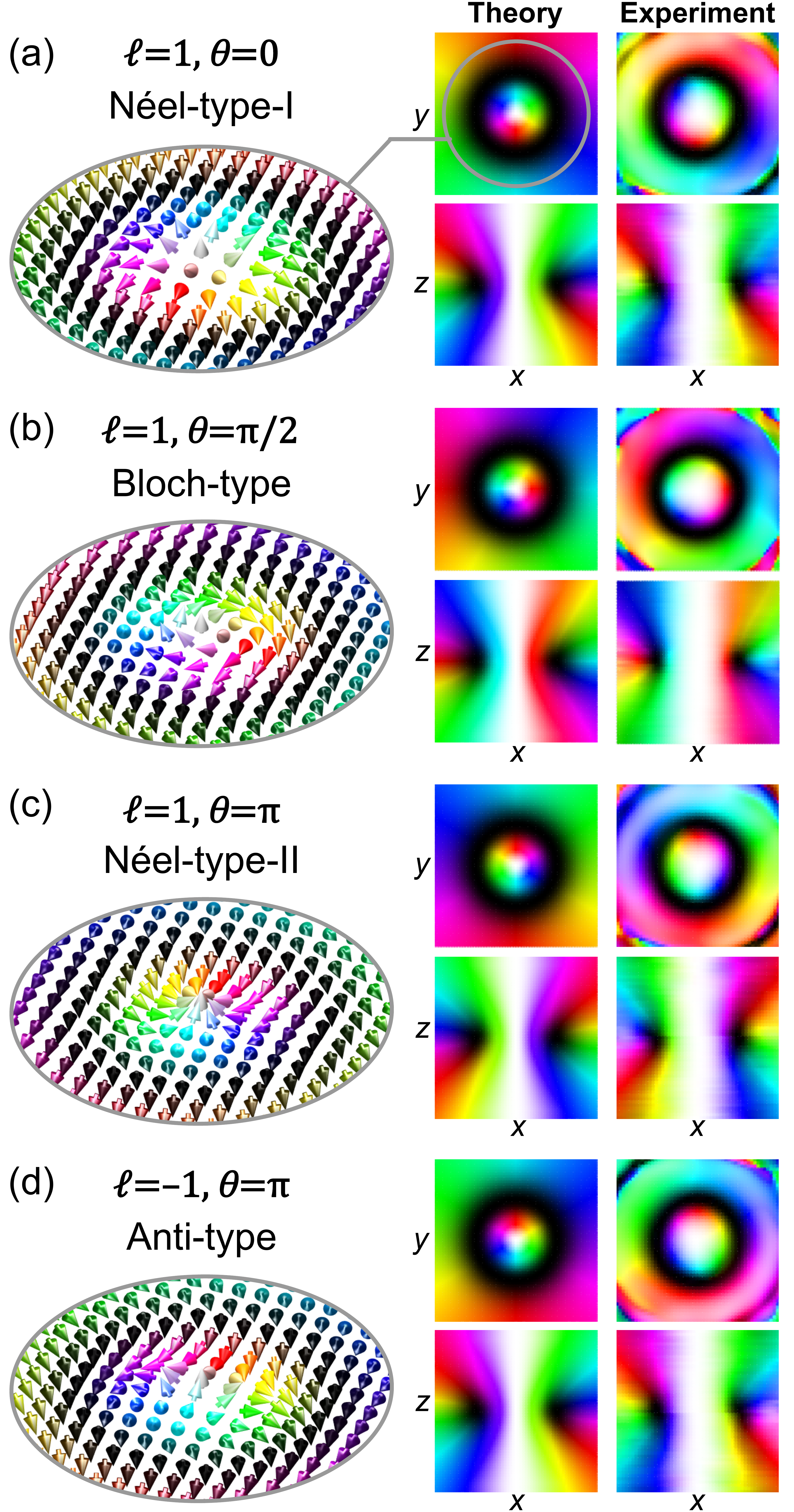}
	\caption{Left: Simulated Stokes vectordistributions in the skyrmionium textures in the $xy$ ($z=0$) plane of the photonic hopfions of (a) N\'eel-type-I ($Q_P=1,Q_V=1,\theta=0$), (b) Bloch-type ($Q_P=1,Q_V=1,\theta=\pi/2$), (c) N\'eel-type-II ($Q_P=1,Q_V=1,\theta=\pi$), and (d) anti-type ($Q_P=1,Q_V=-1,\theta=\pi$). Right: theoretical and experimental polarisation distributions, represented by Poincar\'e parameters (orientation and ellipticity of the polarisation ellipse) in $xy$ and $yz$ planes for the topological hopfions in (a-d). The colour scale is as in Fig.~\ref{f1} (also the same as in Fig.~S4).} 
	\label{f2}
\end{figure}

Using structured light technologies to sculpture complex nonseparable states of an optical vector field~\cite{shen2021rays,shen2022nonseparable,lsa}, a photonic hopfion can be constructed by the polarization Stokes vectors of a 3D structured vector beam with the electric field given by $\bm{\uppsi}(x,y,z)=\psi_R(x,y,z)\hat{\mathbf{e}}_R+\psi_L(x,y,z)\hat{\mathbf{e}}_L$, where $\hat{\mathbf{e}}_R$ ($\hat{\mathbf{e}}_L$) is the eigenstate of right-handed (left-handed) circular polarization and $\psi_R$ ($\psi_L$) is the corresponding spatial mode. The Stokes vector is related to the electric field by $\mathbf{S}=\langle\bm{\uppsi}|\bm{\sigma}|\bm{\uppsi}\rangle$, using the same form of a Hopf map mentioned above. The developed experimental set-up {(see Supplementary Note III for the details of the experiment)} allowed us to generate photonic hopfions with tuneable topology, via the interference of controlled Laguerre-Gaussian (LG) beams, and perform 3D tomography of the Stokes vectors distribution measuring the spatial polarisation distributions. The iso-polarization contours exactly follow the Hopf fibration when $\psi_R=\text{LG}_{0,0}+\text{LG}_{1,0}$ and $\psi_L=\text{LG}_{0,-1}$, where $\text{LG}_{p,\ell}$ is the LG beam with radial index $p$ and azimuthal (orbital angular momentum, OAM) index $\ell$~\cite{sugic2021particle}. Our set-up provides a universal approach to achieve arbitrary photonic hopfions by considering $\psi_R=\text{LG}_{0,0}+\text{e}^{\text{i}\varphi}\text{LG}_{1,0}$ and $\psi_L=\text{e}^{\text{i}\theta}\text{LG}_{0,\ell}$, where $\varphi$ and $\theta$ are two arbitrary intermodal phases, which do not affect the configuration of the Hopf fibration. Photonic hopfions of arbitrary topological numbers $Q_H$ and with all kinds of hopfionic textures can be generated, including N\'eel-, Bloch-, anti-types, and intermediate states. For example, choosing $\ell=1$, the texture of a photonic hopfion can be switched by simply tuning $\theta$ to N\'eel-type-I ($\theta=0$), Bloch-type ($\theta=\pi/2$), N\'eel-type-II ($\theta=\pi$) (Fig.~\ref{f2}(a-c)). Choosing $\ell=-1$, the photonic hopfion of the anti-type is observed  in the case of $\theta=\pi$ (Fig.~\ref{f2}(d)).

The experimentally measured polarisation distributions in the optical hopfions of N\'eel-type-I ($\ell=1,\theta=0$), Bloch-type ($\ell=1,\theta=\pi/2$), N\'eel-type-II ($\ell=1,\theta=\pi$), and Anti-type ($\ell=-1,\theta=\pi$) 
are in excellent agreement to the theoretical results (Fig.~\ref{f2}). 

\begin{figure}[t!]
	\centering
	\includegraphics[width=0.9\linewidth]{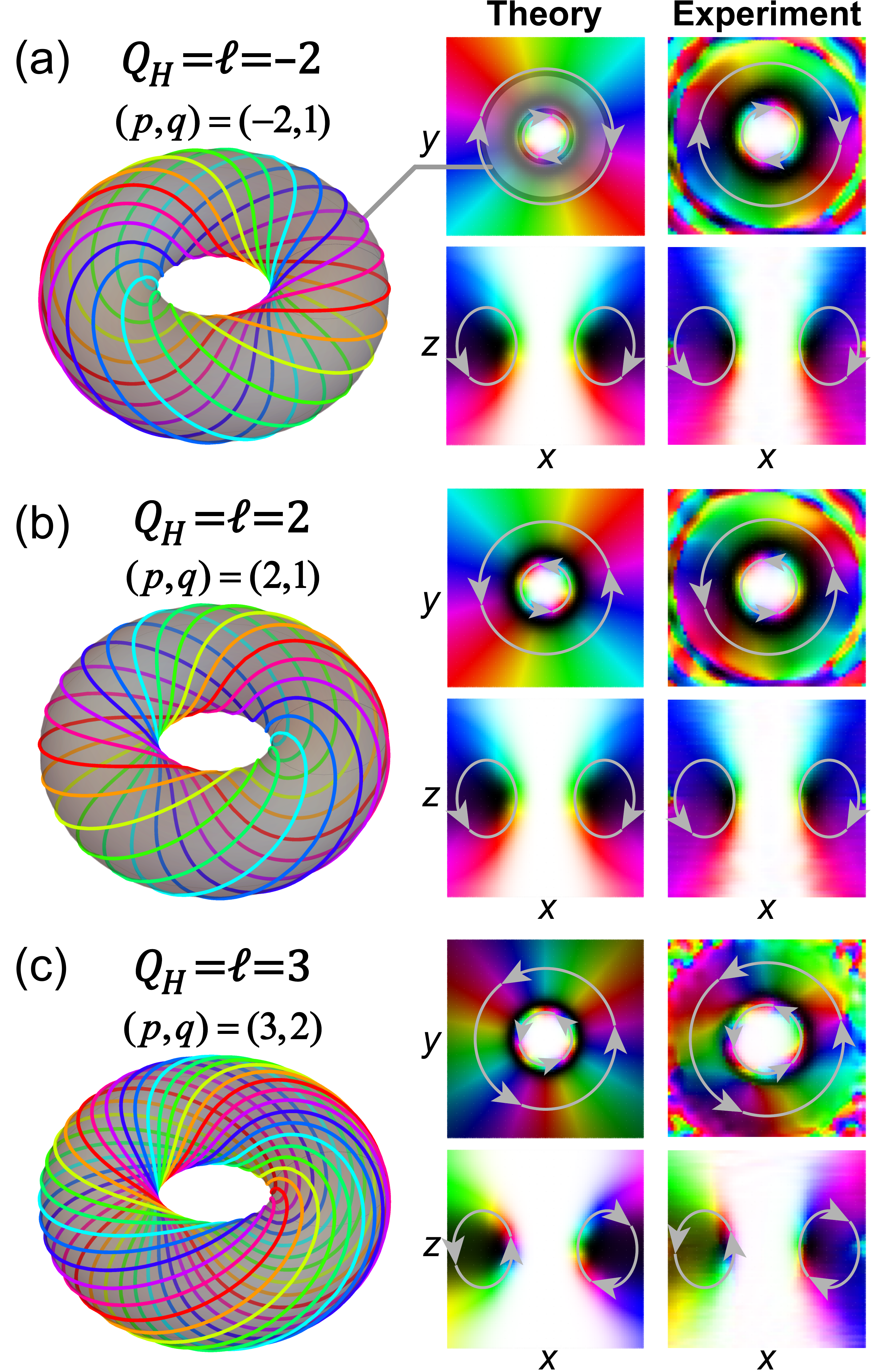}
	\caption{Left: the torus-knot configurations of a toroidal layer in the Hopf fibration for the higher-order hopfions with Hopf indices of  (a) $Q_H=2$, (b) $Q_H=-2$, and (c) $Q_H=-3$. Right: theoretical and experimental polarisation distributions in $xy$ and $yz$ planes for the the hopfions in (a-c). The colour scale is as in Fig. 2.} 
	\label{f3}
\end{figure}

In a fundamental-order hopfion, each fiber as a torus knot in the Hopf fibration goes one time through and one time around the torus, resulting topological charge of $Q_H=\pm1$ (``$\pm$'' decides chirality). For higher-order hopfions, if each torus-knot fiber goes $p$ times through and $q$ times around the torus, the topological charge is $Q_H=\pm p\times q$~\cite{kobayashi2014torus}. The vector fields of various higher-order hopfions can also be characterized by a closed-form expression related to $p$ and $q$ {(see Supplementary Note I)}. Three examples of the torus-knot configurations of three higher-order hopfions with $(p,q)=(-2,1)$, $(p,q)=(2,1)$, $(p,q)=(3,2)$ were generated, using LG beams with $\ell>1$ (Fig.~\ref{f3}). In general, for a given value of $\ell$, the torus-knot indices are determined by $(p,q)=(\ell,|\ell|-1)$. For each higher-order hopfion, the $xy$ cross-section shows corresponding higher-order skyrmionium textures with vorticity equals to $Q_H$, while, the $yz$ cross-section still show two fundamental skyrmions with opposite vorticity.

\begin{figure}[t!]
	\centering
	\includegraphics[width=\linewidth]{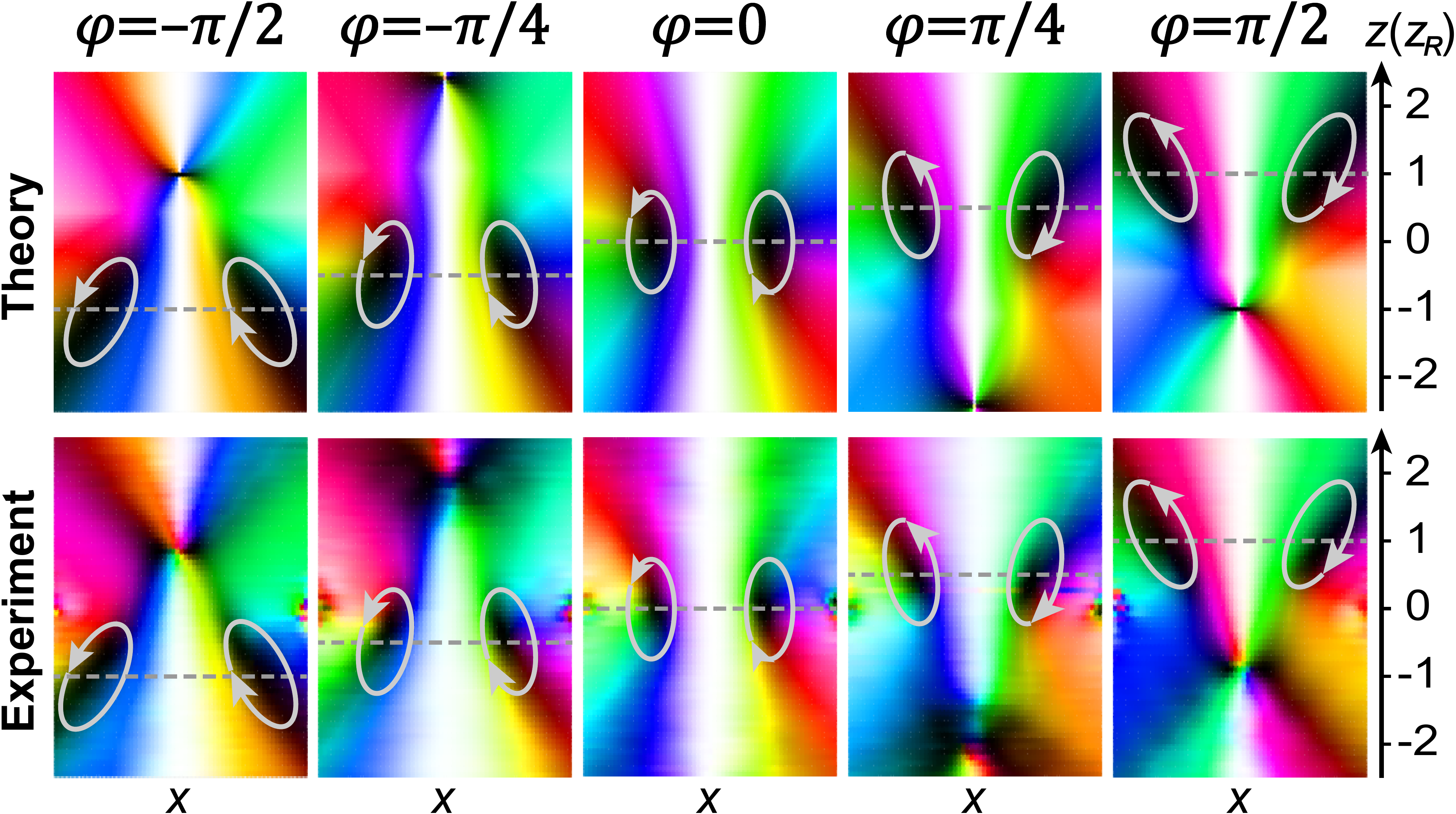}
	\caption{Propagation of photonic hopfions in a free space. Theoretical (top) and experimental (bottom) polarisation distributions in $xz$ plane for different $\varphi$ from $-\pi/2$ to $\pi/2$. The hopfion torus center moves along $z$ from $z=-z_R$ to $z=z_R$. The grey dashed lines mark the location of a hopfion torus centers. Grey arrows indicate the vorticity. The colour scale is as in Fig. 2.}
	\label{f4}
\end{figure}

In the past, a hopfion was always considered as a static topological quasiparticle. Our developed approach provides a unique possibility stimulate and observe free-space transport of hopfions with the preserved topology. This is achieved by tuning the phase parameter $\varphi$. In the initial invetsigations in Figs.~\ref{f2} and~\ref{f3}, $\varphi$=0 and the hopfion center is located at $z=0$. If the phase parameter is tuned, the hopfion configuration controllably propagates along $z$-axis, so that the hopfion centre moves in a $z$ direction. For example, when $\varphi=-\pi/2$, the hopfion is centered at $z=z_R$ (where $z_R$ is Rayleigh length). When the value of $\varphi$ is gradually increased to $\pi/2$, the hopfion center propagates to $z=z_R$ (Fig.~\ref{f4}). This propagation effect can be explained and simulated considering the Gouy phase mediated intramodal phase, resulting in the amplitude profile of $\psi_R$ variation along the radial direction upon propagation~\cite{zhong2021gouy}, which is crucial for determining a 3D polarisation texture of hopfions. As the initial intramodal phase $\varphi$ changed, the center of a hopfion ring, which corresponds to a plane with $\varphi=0$, would move along the $z$-direction (Supplementary Note IV).

In summary, we presented both theoretical and experimental demonstrations of reconfigurable photonic hopfions of higher-orders and their free-space transport. With the development of a high-precision all-digital generation of hopfions, the topological spin textures including N\'eel-, Bloch-, and anti-types can be flexibly manipulated and transformed into each other, demonstrating excellent agreement with the theoretical simulations. The proposed and implemented mechanism to propagate a hopfion topological structure in free space maybe of interest for development of topological classical and quantum information carriers for optical communications and data manipulation.

\bibliographystyle{naturemag}

\textbf{Acknowledgments.} This work was supported, in part, by National Natural Science Foundation of China (grants No. 62075050, 11934013, 61975047), High-Level Talents Project of Heilongjiang Province (2020GSP12), and the European Research Council iCOMM project (789340).

\bibliography{sample}

\end{document}